\documentstyle[aps,prb,multicol,epsf]{revtex}    

\begin{document}            
\newcommand{\eqbreak}{
\end{multicols}
\widetext
\noindent
\rule{.48\linewidth}{.1mm}\rule{.1mm}{.1cm}
}
\newcommand{\eqresume}{
\noindent
\rule{.52\linewidth}{.0mm}\rule[-.1cm]{.1mm}{.1cm}\rule{.48\linewidth}{.1mm}
\begin{multicols}{2}
\narrowtext
\noindent
}

\title{Interaction corrections at intermediate temperatures:  
dephasing time}  
\author{B.N. Narozhny$^{(a)}$, G\'abor Zala$^{(b)}$ 
and I.L. Aleiner$^{(b)}$ } 
\address{ 
$^{(a)}$Physics Department, Brookhaven National Laboratory,  
Upton, NY, 11973-5000, USA \\ 
$^{(b)}$Department of Physics and Astronomy, SUNY at Stony Brook,  
Stony Brook, NY, 11794, USA \\ 
} 
\date{\today}  
\maketitle  
\begin{abstract}                 
 
We calculate the temperature dependence of the weak localization correction 
in a two dimensional system at arbitrary relation between 
temperature, $T$ and the elastic mean free time. We describe the crossover 
in the  dephasing time ${\tau_\varphi(T)}$ 
between the high temperature, $1/\tau_\varphi \simeq  
T^2 \ln T$, and the low temperature $1/\tau_\varphi  \simeq T$ behaviors. 
The prefactors in these dependences are not universal, but are determined by 
the Fermi liquid constant characterising the spin exchange interaction. 
 
\end{abstract}  
\draft  

\begin{multicols}{2}   
 
{\em Introduction} -- The concept of a time scale characteristic of 
electron scattering processes in metals has been the focus of intense 
theoretical research for the last three decades 
\cite{wil,cha,qui,aak,fuk,aar,lrr,cas,fu1,fu2,das,lai,mac,rei,aag,bla}. 
It has been established for disordered systems\cite{aak} that  
the time scale, which corresponds to 
processes that suppress quantum interference (and are thus responsible 
for the temperature dependence of the weak localization correction, 
for example), namely the phase relaxation time $\tau_\varphi$, is 
quite different from the semi-classical time scales, such as 
the energy relaxation time $\tau_E$. 
 
Previous work was mostly focused on the universal  
(independent of interaction strength) contribution of 
the singlet channel interaction both in the diffusive \cite{aak,fuk} 
and ballistic \cite{qui,fuk,das,mac,rei} regimes. The contribution of 
the triplet channel was only considered in the diffusive regime in 
Ref.~\onlinecite{cas}.  In this paper we fill the gaps by considering 
both channels at arbitrary relation between temperature and the 
inverse mean free time and thus describing the cross-over between the 
diffusive and ballistic regimes. We also clarify the relation between 
thus calculated dephasing time and experimentally observable physical 
quantities.

A discussion of phase relaxation should begin by defining a physical 
quantity sensitive to quantum interference  
(since the phase by itself is not an observable quantity). Therefore,  
the precise definition of the phase relaxation time depends on  
the choice of such physical quantity. Consider the weak localization 
correction in two dimensions \cite{aar,aag} in the absence of magnetic field: 
 
\begin{equation} 
\delta\sigma_{WL} (T) = -\frac{e^2}{2\pi^2\hbar }  
\int\limits^\infty_{-\infty} 
\frac{d\epsilon}{4T\cosh^2\frac{\epsilon}{2T}} 
\ln\frac{\tau_\varphi(\epsilon, T)}{\tau}, 
\end{equation} 
\noindent 
where $\tau$ is the transport elastic mean free time. 
With the logarithmic accuracy one can here neglect the dependence of 
$\tau_\varphi$ on $\epsilon$ and estimate \cite{aar,lar} 
 
\begin{equation} 
\delta\sigma_{WL} (T) \approx -\frac{e^2}{2\pi^2\hbar } 
\ln\frac{\tau_\varphi(0, T)}{\tau}.
\end{equation} 
 
\noindent 
In the presence of magnetic field, $H$, perpendicular to the plane of 
the two-dimensional system, the situation is more complicated.  The 
magneto-conductivity can be described \cite{aar,lar,AAA}  
 as 
 
\begin{eqnarray} 
&&\sigma(H,T) - \sigma(0,T) =  
-\frac{e^2}{2\pi^2\hbar}  
\int\limits^\infty_{-\infty} \frac{d\epsilon}{4T\cosh^2\frac{\epsilon}{2T}} 
\label{mc} \nonumber\\
&&\times
\left\{Y[\Omega_H\tau_\varphi(\epsilon, T; H)] 
+\ln \left(\frac{\tau_\varphi(\epsilon, T; H)}{\tau_\varphi(\epsilon, T)}
\right)
\right\}
\end{eqnarray} 
 
\noindent 
where $\Omega_H = 4DeH/\hbar c$, $D$ is the diffusion constant, and  
 
\[ 
Y[x] = \ln\frac{1}{x} - \psi\left(\frac{1}{2} + \frac{1}{x} \right),
\] 
 
\noindent 
with $\psi\left(x \right)$ being the digamma function.
Notice, that the dephasing time now depends on the magnetic field,
see Refs.~\onlinecite{aar,AAA,aag} and Eq.~(\ref{tf}) 
below. We use the notation
\[
\tau_\varphi (\epsilon, T) \equiv \tau_\varphi (\epsilon, T; H=0).
\]

One can simplify the magneto-conductivity (\ref{mc}) using asymptotic 
expressions for $Y[x]$ in order to facilitate comparison with 
experimental data. For strong magnetic fields $\Omega_H\tau_\varphi\gg 
1$ one finds $Y[x\gg1] \approx - \ln x + 2\ln 2+{\bf C}$, 
with ${\bf C}\approx 0.5772\dots$ is the Euler constant.
This yields 
 
\begin{equation} 
\sigma(H,T) - \sigma(0,T) = \frac{e^2}{2\pi^2\hbar} 
\ln \alpha \Omega_H\tau_\varphi^H + {\cal O}\left( 
\frac{1}{\Omega_H\tau_\varphi^H}\right) 
\label{smf} 
\end{equation} 
 
\noindent 
where  $\alpha = 1/(4 e^{\bf C})=0.1403\dots$, and 
 
\begin{equation} 
\tau_\varphi^H = \exp \left( \; \; 
\int\limits^\infty_{-\infty} 
\frac{d\epsilon}{4T\cosh^2\frac{\epsilon}{2T}} 
\ln \tau_\varphi(\epsilon, T) \right) 
\label{th} 
\end{equation} 
 
\noindent 
In the opposite limit of the weak field ($Y[x\ll 1] \approx - x^2/24$) 
the magneto-conductivity is quadratic in $\Omega_H$ 
 
\begin{equation} 
\sigma(H,T) - \sigma(0,T) = \frac{e^2}{48\pi^2\hbar} 
\left( \Omega_H\tau_\varphi^T \right)^2 
\label{laf} 
\end{equation} 
 
\noindent 
where 
 
\begin{equation} 
\tau_\varphi^T = 
\left[ 
\int\limits^\infty_{-\infty} 
\frac{d\epsilon}{4T\cosh^2\frac{\epsilon}{2T}} 
\tau_\varphi^2(\epsilon, T) 
\right]^{1/2}. 
\label{tt} 
\end{equation} 
 
\noindent 
As we show below, the two scales $\tau_\varphi^T$ and $\tau_\varphi^H$ 
may differ by a numerical factor which is important for 
a quantitative comparison with experiments.

As a function of $\epsilon$, $T$ and $H$ the phase relaxation 
time is given by \cite{aar} 
 
\begin{eqnarray} 
&&\frac{1}{\tau_{\varphi}(\epsilon,T;H)}= \int 
\limits^\infty_{{\rm max}[\Omega_H,\tau_{\varphi}^{-1}]} 
d\omega \; \omega \; A(\omega) 
\label{tf}\\ 
&& \quad \times\Bigg[ \coth\frac{\omega}{2T}  
- \frac{1}{2} 
\left(\tanh\frac{ \omega-\epsilon}{2T}  
+\tanh\frac{\omega+\epsilon  }{2T} \right) 
\Bigg].\nonumber 
\end{eqnarray} 
 
\noindent 
Kernel $A(\omega)$ contains all the 
information about matrix elements of the interaction and 
wavefunctions of the disordered system, see e.g. \cite{aar,aag}, 
and this is precisely the same kernel which enters into 
 the inelastic collision integral in the Boltzmann equation, 
see e.g. Ref.~\onlinecite{aar,bla}: 
 
\begin{eqnarray} 
\label{st} 
\mbox{St}_{in}&&\left\{f(\epsilon)\right\} = 
\int d\omega \int d \epsilon_1 A(\omega)  
f( \epsilon_1)  
\left[1 -  f(\epsilon_1 - \omega)\right] 
\label{inelastic} 
\\ 
&& 
\nonumber\\ 
&& 
\times  
\Big\{ - f(\epsilon)  
\left[1 -  f( \epsilon + \omega)\right] 
+ \left[1 - f( \epsilon)\right]f( \epsilon - \omega) 
\Big\}, 
\nonumber 
\end{eqnarray} 
 
\noindent 
where $f$ is the distribution function of the quasiparticles.

{\em Calculation of the kernel} -- The kernel $A(\omega)$ can be 
obtained from the quantum kinetic equation (see Ref.~\onlinecite{us1}) 
and it is expressed in terms of the interaction propagator ${\cal D}^R$ 
and the propagators $\langle D \rangle$ describing semi-classical 
dynamics of non-interacting electrons 
 
\begin{eqnarray} 
A(\omega)&& = \frac{2\nu}{\pi} \int \frac{d^2q}{(2\pi)^2} \; 
\Big( [{\bf Re} \; \langle D \rangle ]^2 |{\cal D}^R_S(\omega, q)|^2  
\nonumber\\ 
&& 
\nonumber\\ 
&& 
+ {\bf Tr} [{\bf Re} \; \langle\widehat{ D }\rangle]
{\cal \widehat D}^R_T(\omega, q) [{\bf Re} \; \langle\widehat{ D }\rangle ]
[{\cal \widehat D}^R_T(\omega, q)]^* \Big). 
\label{ao} 
\end{eqnarray} 
 
\noindent
In the absence of external magnetic field and spin-orbit interaction the
semi-classical propagator is diagonal 
$\langle\widehat D \rangle = \delta_{ij}\langle D \rangle$ and
 
\begin{equation} 
\langle D \rangle = \frac{1}{\sqrt{(-i\omega +1/\tau)^2 + v_F^2 q^2} - 1/\tau}. 
\label{d} 
\end{equation} 
 
\noindent 
It becomes the usual diffuson \cite{aar} in the diffusive limit 
$\omega, v_F q \ll 1/\tau$.

The interaction propagator in the singlet channel is given in terms of 
$\langle D \rangle$ and the Fermi liquid constant $F_0^\rho$ \cite{us1,lfl}

\begin{equation} 
{\cal D}^R_S(\omega, q) = -\frac{1}{\nu}
\frac{\nu V_0(q)+F_0^\rho}{1+(\nu V_0(q)+F_0^\rho)   
[1+i\omega\langle D \rangle]}.
\label{sp} 
\end{equation} 

\noindent
The interaction propagator in the triplet 
channel is a diagonal matrix
in spin indices:
 
\begin{equation} 
{\cal \widehat D}^R_T(\omega, q) = - \frac{\delta_{ij}}{\nu}
\frac{F_0^\sigma}{1 +F_0^\sigma   
[1+i\omega\langle D \rangle]}, 
\label{pr} 
\end{equation} 
 
\noindent 
where $F_0^\sigma$ is the Fermi-liquid constant in the triplet channel 
\cite{us1,lfl}. At distances larger that the screening radius one can take
the unitary limit in Eq.~(\ref{sp}), $\nu V_0(q)\to \infty$ 
which coinsides with the  
$F_0^\sigma\rightarrow\infty$ limit in 
Eq.~(\ref{pr}) [see Ref.~\onlinecite{us1} for more details]. 
 
We now evaluate the triplet contribution to the kernel $A(\omega)$, 
separating Eq.~(\ref{ao}) into the sum 
 
\begin{equation}
A(\omega) = 3 A_T(\omega) + A_S(\omega). 
\label{asum}
\end{equation} 
 
\noindent 
Using Eqs.~(\ref{pr}) and (\ref{d}) one can rewrite Eq.~(\ref{ao}) as 
 
\begin{equation} 
A_T(\omega) = \frac{1}{\pi\omega}\; {\bf Im} \;\int \frac{d^2q}{(2\pi)^2}  
{\cal D}^R_T(\omega, q)  
\left[ \langle D \rangle + \langle D \rangle^*\right] 
\label{a} 
\end{equation} 
 
\noindent 
The kernel $A_S(\omega)$ can be obtained from Eq.~(\ref{a}) by taking the 
limit $F_0^\sigma\rightarrow\infty$. 
The momentum integral in Eq.~(\ref{a}) diverges logarithmically in the 
ultraviolet.  This divergence is typically \cite{fuk,qui,rei} cut off 
at a scale of order $k_F$ (the precise definition of such cut off is 
important only for the numerical factor under the logarithm which we 
believe is beyond the accuracy of any discussion). In the diffusive 
limit the divergence does not appear since one is limited by small 
momenta $v_Fq\ll 1/\tau$. It is therefore convenient to separate 
$A_T(\omega)$ into two parts, roughly corresponding to the ballistic 
and diffusive asymptotics, respectively: 
 
\begin{equation} 
A_T(\omega) = A_1(\omega) + A_2(\omega). 
\label{ar} 
\end{equation} 
 
The ``ballistic'' term appears from integrating over momenta larger than 
inverse elastic mean free time and/or frequency. With logarithmic 
accuracy we find (here the product of the Fermi momentum and 
the Fermi velocity [renormalized by interaction] 
is denoted by $E_F=v_F k_F /2$) 
 
\begin{mathletters} 
\begin{eqnarray} 
&&A_1(\omega) =\frac{(F_0^\sigma)^2}{4\pi E_F (1+F_0^\sigma)^2} 
\ln\frac{E_F^2} 
{\tau^{-2} + b(F_0^\sigma)\omega^2}, 
\label{a1} \\
&&b(x) \approx \frac{1+x^2}{(1+x)^2}
\nonumber
\end{eqnarray} 
 
\noindent 
This term has the logarithmic frequency dependence similar to the one 
obtained by several authors for the singlet channel interaction 
\cite{fuk,rei}.  
 
The ``diffusive'' term, i.e. coming from integrating over small 
momenta, is a generalization of the standard result \cite{aar}.  For 
simplicity, we show the kernel $A_2(\omega)$ in the two limiting 
cases. For small energy transfers $\omega\tau\ll 1$ we find

\begin{eqnarray} 
A_2(&&\omega\tau\ll 1) = \frac{F_0^\sigma}{(1+F_0^\sigma)(2+F_0^\sigma)} 
\frac{1}{2 g \omega} 
\nonumber\\ 
&& 
\nonumber\\ 
&& 
\times 
\Big[ 
(1+F_0^\sigma)\arctan\frac{1}{\omega\tau} 
- 
\arctan\frac{1+F_0^\sigma}{\omega\tau} 
\Big]. 
\label{a2d} 
\end{eqnarray}

\noindent 
where $g=2\pi\hbar/e^2 R_\Box$ is the dimensionless conductance of the 
system, and  $R_\Box$ is the sheet resistance. 
 
Note, that this result reduces to that of the diffusive theory for 
$\omega\tau\ll 1+F_0^\sigma$ (see Refs.~\onlinecite{us1} and 
\onlinecite{hal} for a more detailed discussion of the crossover to 
the ballistic regime). For the singlet channel (which can be obtained 
by setting $F_0^\sigma\rightarrow\infty$) Eq.~(\ref{a2d}) coincides 
with the standard result \cite{aar}.

For large energy transfers $\omega\tau\gg 1$ integration over small 
moments yields a correction to Eq.~(\ref{a1}): 
 
\begin{eqnarray} 
A_2(&&\omega\tau\gg 1) = \frac{1}{1+F_0^\sigma}\frac{1}{2 g \omega} 
\nonumber\\ 
&& 
\nonumber\\ 
&& 
\times 
\Big[\arctan\frac{1+F_0^\sigma}{\omega\tau(1-F_0^\sigma)} 
+(1+F_0^\sigma)\arctan\frac{1}{\omega\tau}\Big]. 
\label{a2b} 
\end{eqnarray} 
\label{ker} 
\end{mathletters} 
 
\noindent 
For numerical reasons, contributions of Eq.~(\ref{a2b}) to the final 
results are small compared with that of 
Eq.~(\ref{a1}) for all temperature regimes.

{\em Results for the dephasing time} -- We now use the explicit form 
of the kernel Eqs.~(\ref{ker}) to find the dephasing time from the 
self-consistency equation (\ref{tf}).  According to Eqs.~(\ref{asum}) and 
(\ref{ar}), it takes the form 
 
\begin{equation} 
\frac{1}{\tau_\varphi(\epsilon,T; H)} 
 = I_1\left(\epsilon,T\right) + I_2 
\left(T,{\rm max}[\Omega_H,\tau_{\varphi}^{-1}]\right). 
\label{i12} 
\end{equation} 
 
\noindent 
The kernel (\ref{a1}) is not diverging at small energy transfers, and 
the calculation of term $I_1$ can be performed by setting the lower 
limit of integration in Eq.~(\ref{tf}) to zero.  We find with 
the logarithmical accuracy 
 
\begin{mathletters} 
\label{i} 
\begin{eqnarray} 
&&I_1 =\frac{\pi^2T^2+\epsilon^2}{8\pi E_F} \label{i1} \\
&&
\times\left[\frac{3(F_0^\sigma)^2}{(1+F_0^\sigma)^2}  
\ln\left( 
\frac{E_F^2}{b(F_0^\sigma)
T^2+\tau^{-2}}\right)
+\ln\left( 
\frac{E_F^2}{
T^2+\tau^{-2}}\right)
\right], 
\nonumber
\end{eqnarray} 
where function $b(x)$ is defined in Eq.~(\ref{a1}).
 
The diffusive term $I_2$ is logarithmically divergent at the lower 
limit. The divergent contribution is independent of $\epsilon$ and the 
divergence is cut according to Eq.~(\ref{tf}) 
 
\begin{equation} 
I_2\left(T,\Omega\right) =  
\left(1+\frac{3(F_0^\sigma)^2}{(1+F_0^\sigma)(2+F_0^\sigma)}\right) 
\frac{T}{g}\ln \left(\frac{T}{\Omega}\right). 
\label{i2} 
\end{equation} 
\end{mathletters}

1. Diffusive limit ($T\tau\ll 1+F_0^\sigma$).  
In this case the dephasing time follows from the self-consistency  
Eq.~(\ref{tf}) where one has to include the 
singlet channel contribution. The time scales $\tau_\varphi^{T,H}$ 
defined 
in Eq.~(\ref{smf}) are given by 
 
\eqbreak

\begin{eqnarray} 
\frac{1}{\tau_\varphi^T} =  \frac{1}{\tau_\varphi^H} =  
\left(1+\frac{3(F_0^\sigma)^2}{(1+F_0^\sigma)(2+F_0^\sigma)}\right) 
\frac{T}{g}\ln \Big[g(1+F_0^\sigma)\Big] 
+ \frac{\pi}{4}\left(1+\frac{3(F_0^\sigma)^2}{(1+F_0^\sigma)^2}\right) 
\frac{T^2}{E_F}\ln(E_F\tau). 
\label{dif} 
\label{dd}
\end{eqnarray}

\eqresume

\noindent 
At stronger fields $\Omega_H > T$ equation (\ref{dd}) is not  
applicable, since in that regime the interaction correction in the 
Cooper channel becomes of the same order as the weak localization  
correction \cite{aar}. We will not dwell on this issue here. 
 
The result (\ref{dd}) are valid while the condition 
$g(1+F_0^\sigma)\gg 1$ holds. The same condition guarantees the  
exponential smallness of temperature independent dephasing induced 
by spontaneously spin polarized regions \cite{nal}. 
 
Comparing the dominant, diffusive term [the first term in 
Eq.~(\ref{dif}), which came from integrating Eq.~(\ref{a2d})] to the 
second, ballistic term, we find that the two become of the same order 
when $T\tau\sim 1+F_0^\sigma$, which sets the limit of applicability 
to the purely diffusive theory.

2. Ballistic limit. At temperatures $T\tau\gg 1+F_0^\sigma$ 
the leading asymptotics is controlled by Eq.~(\ref{i1}). 
In this regime we can no longer neglect the $\epsilon$ dependence in 
$\tau_\varphi(\epsilon, T)$. As a result, the time scales 
$\tau_\varphi^T$, see  
Eq.~(\ref{tt}), and $\tau_\varphi^H$, see Eq.~(\ref{th}), 
are different by a numerical factor: 
 
\begin{mathletters} 
\begin{eqnarray} 
\tau_\varphi^{H, T} = \tau_\varphi (0, T) B^{H, T} 
\label{rb1} 
\end{eqnarray} 
 
\noindent 
where  
 
\begin{equation} 
\frac{1}{\tau_\varphi(0, T)}
=\frac{\pi T^2}{4 E_F} 
\Bigg[\ln\frac{E_F}{T}+ 
\frac{3(F_0^\sigma)^2}{(1+F_0^\sigma)^2} 
\ln\frac{E_F}{T \sqrt{b(F_0^\sigma)}}\Bigg], 
\label{rb0} 
\end{equation} 
 
\noindent 
where $b(x)$ is defined in Eq.~(\ref{a1}),
and the numerical factors are 
 
\begin{eqnarray*} 
B^T = \left[\int\limits_{0}^\infty  
\frac{dz}{\cosh^2z}\left(1+\frac{4z^2}{\pi^2}\right)^{-2} 
\right]^{1/2}\approx 0.8437\dots; 
\end{eqnarray*} 
 
\begin{eqnarray*} 
B^H=\exp\left[-\int\limits_{0}^\infty  
\frac{dz}{\cosh^2z} 
\ln \left(1+\frac{4z^2}{\pi^2}\right)\right]\approx 0.7931\dots. 
\end{eqnarray*} 
 
\label{br} 
\end{mathletters} 
 
The observable 
dephasing times are different from $\tau_\varphi(0, T)$ by the above 
numerical constants. 
The temperature dependence of $\tau_\varphi(0, T)$ 
Eq.~(\ref{rb0}) coincides with that found earlier in 
Refs.~\onlinecite{qui,fuk,das,mac,rei} for the case $F_0^\sigma =0$ 
\cite{Footnote1,Footnote2}.  
The numerical factor $\pi/4$ 
is the same as in Refs.~\onlinecite{das,mac,rei}, while 
Ref.~\onlinecite{fuk} reports the factor $\pi/2$ and 
Ref.~\onlinecite{qui} reports the factor of $1/2\pi$.  
The correct numerical factor $\pi/4$ was also obtained in 
Ref.~\onlinecite{lai}, however it was claimed that this result should
be further renormalized (reduced by a factor of $4$) by taking into 
account higher order forward
scattering processes. We believe such renormalization is erroneos and
is a result of misidentification of the Fermi liquid constant in the
singlet channel \cite{la1}.

{\em Summary} -- We have calculated the temperature dependence of the  
weak localization correction at arbitrary relation between $T$ and 
elastic mean free time $\tau$. The prefactors in  
the temperature dependencies of the 
dephasing rate are not universal and are determined by the single 
Fermi liquid constant $F_0^\sigma$. The very same constant determines  
the temperature dependence of the longitudinal resistivity \cite{us1}, 
the Hall coefficient \cite{hal}, and magnetoresistance in the  
parallel magnetic field \cite{us3}. Because the number of different 
observable quantities exceeds the number of input parameters this  
theory posesses the predictive power.

{\em Acknowledgments} -- We are grateful to M.Yu. Reizer and B. Laikhtman 
for interesting discussions, and to A.D. Mirlin for pointing an error
in an earlier version of this paper.
One of us (I.A.) was supported by the Packard 
foundation. Work at BNL is supported by the US DOE under Contract No 
DE-AC02-98CH10886.

\end{multicols}  
  

\begin{references} 
  
\bibitem{wil} C. Hodges, H. Smith, and J.W. Wilkins,  
Phys. Rev. B {\bf 4}, 302 (1971). 
\bibitem{cha} A.V. Chaplik, Zh. Eksp. Teor. Fiz. {\bf 60}, 1845 (1971)  
[Sov. Phys. JETP {\bf 33}, 997 (1971)]. 
\bibitem{qui} G.F. Giuliani and J.J. Quinn, Phys. Rev. B {\bf 26}, 4421 (1982). 
\bibitem{aak} B.L. Altshuler, A.G. Aronov, and D.E. Khmelnitsky,  
J. Phys. C {\bf 15}, 7367 (1982). 
\bibitem{fuk} H. Fukuyama and E. Abrahams, Phys. Rev. B {\bf 27}, 5976 (1983). 
\bibitem{aar} B.L. Altshuler and A.G. Aronov in  
{\em Electron-Electron Interactions in Disordered Systems},   
eds. A.L. Efros, M. Pollak (North-Holland, Amsterdam, 1985).  
\bibitem{lrr} P.A. Lee and Ramakrishnan, Rev. Mod. Phys. {\bf 57}, 287 (1985). 
\bibitem{cas} C. Castellani, C. DiCastro, G. Kotliar, and P.A. Lee, 
Phys. Rev. Lett. {\bf 56}, 1179 (1986). 
\bibitem{fu1} H. Fukuyama and Y. Hasegawa, Prog. Theor. Phys. Suppl.
{\bf 101}, 441 (1990).
\bibitem{fu2} H. Fukuyama, Y. Hasegawa, and O. Narikiyo, J. Phys. Soc. Jpn.
{\bf 60}, 2013 (1991).
\bibitem{das} L. Zheng and S. Das Sarma, Phys. Rev. B {\bf 53}, 9964 (1996). 
\bibitem{lai} D. Menashe and B. Laikhtman, Phys. Rev. B {\bf 54}, 11561 (1996). 

\bibitem{mac} T. Jungwirth and A.H. MacDonald,  
Phys. Rev. B {\bf 53}, 7403 (1996). 
\bibitem{rei} M. Reizer and J.W. Wilkins, Phys. Rev. B {\bf 55}, R7363 (1997). 
\bibitem{aag} I.L. Aleiner, B.L. Altshuler, and M.E. Gershenson, 
Waves Random Media {\bf 9}, 201 (1999). 
\bibitem{bla} I.L. Aleiner and Ya.M. Blanter, 
Phys. Rev. B {\bf 65}, 115317 (2002). 
\bibitem{lar}  
B.L. Altshuler, D. Khmelnitzkii, A.I. Larkin, and P.A. 
Lee, Phys. Rev. B {\bf 22}, 5142 (1980); 
S. Hikami, A.I. Larkin, and Y. Nagaoka, 
Prog. Theor. Phys. {\bf 63}, 707 (1980).  

\bibitem{AAA} E.L. Altshuler, B.L. Altshuler, and A.G. Aronov,
Solid State Commun., {\bf 54}, 617 (1985). 

\bibitem{us1} G\'abor Zala, B.N. Narozhny, and I.L. Aleiner,  
Phys. Rev. B {\bf 64}, 214204 (2001). 
\bibitem{lfl} L.D. Landau, Zh. Eksp. Teor. Fiz. {\bf 30}, 1058 (1956);  
{\it ibid} {\bf 32}, 59 (1957).  
\bibitem{hal} G\'abor Zala, B.N. Narozhny, and I.L. Aleiner,  
Phys. Rev. B {\bf 64}, 201201 (2001). 
\bibitem{nal} B.N. Narozhny, I.L. Aleiner, and A.I. Larkin,  
Phys. Rev. B {\bf 62}, 14898 (2000). 
\bibitem{Footnote1} We note that the expression of the 
relaxation rate in terms of the Fermi liquid constant is 
somewhat similar to 
the taking into account of so called ``local field corrections'' of 
Ref.~\onlinecite{mac}. We believe, that the deviation of our results  
from Eq.~(24) of this reference is caused by the fact that the singlet 
(charge) and triplet, $L_z=0$, contributions were not separated 
properly [see Eq.~(19) of this reference]. 
\bibitem{Footnote2} Reference \onlinecite{rei} argues that the 
contribution of other ``non-golden rule'' diagrams, see Fig.~1 of 
Ref.~\onlinecite{rei} decreases the coefficient by a factor of 2. We think 
that this statement is not correct -- this diagram, in fact, describes 
the renormalization of
$F_0^\rho$ and 
the effect of interaction in the Cooper channel on inelastic 
processes. The latter is suppresed by the factor of $1/\ln^2(E_F/T)$ 
for the repulsive interaction. 
\bibitem{la1} The correct form of the propagator in the singlet channel is
given by Eq.~(\ref{sp}), which at distances greater than the screening
radius reduces to ${\cal D}^R_S(q=0) = 1/\nu$. We believe
that the resummation of Ref.~\onlinecite{lai} results in the incorrect form 
of the singlet propagator,
${\cal D}^R_S(q) = F_0^\rho/\nu + V(q)/[1+\nu V(q)]$, where for forward 
scattering
$F_0^\rho\rightarrow -1/2$, hence the extra 
factor of $1/4$ in the final result 
of Ref.~\onlinecite{lai}. Opinion of B. Laikhtman (private communication) is
that disagreement by a factor of four is caused by algebraic mistakes
in Eqs. (24) and (25) of Ref.~\onlinecite{lai}.
\bibitem{us3} G\'abor Zala, B.N. Narozhny, and I.L. Aleiner, 
Phys. Rev. B, {\bf 65}, 020201(R), (2002). 
 
\end{references}
\end{document}